\documentclass[trackchanges]{aastex701}

\usepackage{amssymb, amsmath}

\newcommand\offset{$175 \ \mathrm{pc}$}

\newcommand\sourcefullname{MG~B2016+112}
\newcommand\sourceshortname{MG~B2016}
\newcommand\sourceredshift{$3.273$}

\newcommand{\maxsephst}{$3\farcs75$} 
\newcommand{\ztproximateabs}{$3.27$}

\newcommand{\hst}{\mbox{\em HST\/}}

\newcommand{\axaf}{\mbox{\em Chandra\/}}

\newcommand{\asca}{\mbox{\em ASCA\/}}

\newcommand{\cfht}{\mbox{\em CFHT\/}}

\newcommand{\massalowlim}{$5.5\times 10^6 M_\odot$}
\newcommand{\massblowlim}{$6.3\times 10^6 M_\odot$}
\newcommand{\massclowlim}{$4.8\times 10^6 M_\odot$}

\newcommand{\lyalpha}{\mbox{Ly$\alpha$\/}}
\newcommand{\halpha}{\mbox{H$\alpha$\/}}
\newcommand{\hbeta}{\mbox{H$\beta$\/}}
\newcommand{\carbonfour}{\mbox{\textnormal{C\,{\sc iv}}\/}}
\newcommand{\carbonthree}{\mbox{\textnormal{C\,{\sc iii]}}\/}}
\newcommand{\heliumtwo}{\mbox{\textnormal{He\,{\sc ii}}\/}}

\newcommand{\nitrogenfive}{\mbox{\textnormal{N\,{\sc v}}\/}}
\newcommand{\magnesiumone}{\mbox{\textnormal{Mg~{\sc i}}\/}}

\newcommand{\muse}{\mbox{\em MUSE\/}}
\newcommand{\jwst}{\mbox{\em JWST\/}}

\usepackage{subcaption}

\usepackage{xcolor}

\begin{document}

\title{\muse\ spectroscopy of the compact dual AGN in the $z=$~\sourceredshift\ radio-loud gravitational lens \sourcefullname.}

\author[orcid=0009-0004-0648-9529]{Jianghao Huyan}
\altaffiliation{These authors contributed equally to this work.}
\affiliation{Current address: School of Astronomy and Space Science, Nanjing University, Nanjing, Jiangsu 210093, People’s Republic of China}
\affiliation{Department of Physics \& Astronomy, University of South Carolina, Columbia, SC 29208, USA}
\email[show]{jhuyan@email.sc.edu}  

\author[orcid=0000-0003-3814-6796]{J\'ulia M.~Sisk-Reyn\'es}
\altaffiliation{These authors contributed equally to this work.}
\affiliation{Center for Astrophysics $|$ Harvard \& Smithsonian, 60 Garden Street, Cambridge, MA 02138, USA}
\email[show]{jsiskre1@umbc.edu}  
\affiliation{Department of Physics, University of Maryland Baltimore County, 1000 Hilltop Cir, Baltimore, MD 21250, USA}  

\author[orcid=0000-0001-8252-4753]{Daniel A.~Schwartz} 
\affiliation{Center for Astrophysics $|$ Harvard \& Smithsonian, 60 Garden Street, Cambridge, MA 02138, USA}
\email[hide]{das@cfa.harvard.edu}

\author[orcid=0000-0002-2587-2847]{Varsha P.~Kulkarni}
\affiliation{Department of Physics \& Astronomy,
University of South Carolina,
Columbia, SC 29208, USA}
\email[hide]{kulkarni@sc.edu}  

\author[0000-0001-5655-4158]{Anna Barnacka}
\affiliation{Center for Astrophysics $|$ Harvard \& Smithsonian, 60 Garden Street, Cambridge, MA 02138, USA}
\email[hide]{abarnacka@cfa.harvard.edu}

\author[0000-0002-1616-1701]{Adi Foord}
\affiliation{Department of Physics, University of Maryland Baltimore County, 1000 Hilltop Cir, Baltimore, MD 21250, USA} 
\email[hide]{foord@umbc.edu}

\author[orcid=0009-0000-9392-3557,gname='Charlotte', sname='Eades']{Charlotte A.~Eades}
\email[hide]{cae1g22@soton.ac.uk} 
\affiliation{School of Physics and Astronomy, University of Southampton, Southampton, SO17 1BJ, UK}
\affiliation{Department of Physics, Astrophysics, University of Oxford, Denys Wilkinson Building, Keble Road, Oxford
OX1 3RH, UK}

\begin{abstract}

Dual active galactic nuclei (AGN) are unique laboratories for studying how galaxy mergers trigger accretion onto supermassive black holes (SMBHs).~However, only a few dual AGN with projected separations~$\lesssim 1 \ \mathrm{kpc}$ at redshifts $z>1$~have been confirmed to date.~We present archival \muse\ spectroscopy of the three resolved lensed images of the radio-loud type-II system \sourcefullname.~At $z=$~\sourceredshift~and at a projected separation of \offset,~this is the most compact confirmed dual AGN at $z > 1$.~We detect \lyalpha, \carbonfour, \nitrogenfive, \heliumtwo, and \carbonthree\ emission across images A--C after correcting for ISM absorption at $z \sim 0$. We find that the \lyalpha, \carbonfour, and \nitrogenfive\ lines are affected by saturated absorption at $z\sim$~\ztproximateabs.~The \heliumtwo/\carbonthree\ emission-line ratios are consistent between images A and B but differ significantly from image C.~We fit the \lyalpha\ and \carbonfour\ emission regions in images A--C with a broad and a narrow Gaussian and use the broad Gaussian components in \carbonfour\ to place lower limits on their respective SMBH masses:~$\gtrsim$~\massalowlim,~$\gtrsim$~\massblowlim,~and~$\gtrsim$~\massclowlim.~We then combine these SMBH mass limits with the magnification-corrected luminosities derived from both the \muse\ data~and recent \axaf\ X-ray observations to estimate their Eddington ratios.~Our findings motivate deep, spatially resolved near-IR spectroscopy to independently determine the SMBH masses from \halpha\ and \hbeta\ emission and determine the physical conditions of the host galaxy.

\end{abstract}

\keywords{\uat{Active galactic nuclei}{16} --- \uat{Supermassive black holes}{1663} -- \uat{Strong gravitational lensing}{1643} -- \uat{Emission line galaxies}{459} -- \uat{High-redshift galaxies}{734} -- -- \uat{Galaxy mergers}{608} }

\section{Introduction} 
\label{sec:sec1_intro}
Theoretical models of cosmic structure formation predict that galaxy mergers play an important role in the late stages of galaxy evolution, likely triggering enhanced accretion onto supermassive black holes \citep[SMBHs;][]{diMatteo_2005_cosmo,hopkins_2005_cosmo,springel_2005_cosmo}.~Observational efforts have therefore sought to identify and characterize dual\setcounter{footnote}{0}\footnote{Here, we use the term ``dual'' to refer to a SMBH pair where the two SMBHs are not yet in a ``binary'' phase as they may (or may not) evolve into a gravitationally bound system.} active galactic nuclei (AGN) across a broad range of physical separations and cosmic timescales \citep[e.g.][]{Derosa2019,comerford2013} using low-frequency radio imaging \citep{damato_lotts_duals} and radio very-long-baseline interferometry \citep[VLBI;][]{burke-spolaor_2011_radio-census-duals}, optical imaging and spectroscopy \citep{liu_i_2010_duals,liu_ii_2010_duals,comerford_2012_dual-searches,mannucci2022_gmp,vodka_iii_2023,mannucci2023_gmp,vodka_ii_2024,vodka_i_2025}, and X-ray observations \citep{komossa_2003_ngc6240,koss2012,foord_2020_dual-candidates-baymax,derosa2023,Battistini_2026_duals}.

The most compact confirmed dual AGN known at low redshift has a projected separation of $\sim230$~pc \citep[at $z=0.03$;][]{koss_2023_dual-confirmation}, while the tightest-separation dual AGN candidate has a projected separation of $\sim 100 \ \mathrm{pc}$ \citep[at $z=0.016$;][]{trindade-falcao_2024_dualAGNcand,lamperti_2026}.~In contrast, at $z>1$, most compact dual AGN candidates identified through optical varstrometry have projected separations $\gtrsim 1 \ \mathrm{kpc}$~\citep{chen_2022_varstrometry}.~Spatially resolved spectroscopy of distinct broad-line regions (BLRs) with the \textit{James Webb Space Telescope}~(\jwst)~has recently revealed multiple SMBHs residing in merging galaxies at even higher redshifts ($z \sim 5)$, with projected separations between the SMBHs of $ \sim \mathrm{few} \times  100 \ \mathrm{pc}- 1 \ \mathrm{kpc}$~\citep[e.g.~see][and references therein]{ubler_2026_blackthunder}. In general, identifying and probing sub-kpc AGN pairs at $z\sim 1$ remains challenging when their projected separation is below the resolution of direct imaging~\citep[see discussions in][]{sisk-reynes_2026_gralj0659}. Strong gravitational lensing offers a unique solution by producing multiple, flux-magnified images of individual, distant sources, enabling fine-scale source-plane reconstruction through detailed lens modeling~\citep{barnacka_2017_rev,barnacka_2018_rev}.

\sourcefullname~(\sourceshortname\ herein) provides a rare opportunity to probe a compact dual AGN at $z=$~\sourceredshift~through strong gravitational lensing.~\sourceshortname\ is lensed into three distinct images (A--C; Figure \ref{fig:figure1_hst+muse}), each of which is further resolved into multiple VLBI sub-components \citep{more2009,Spingola2019}.~Early optical spectroscopy of images A and B established their association with an AGN by revealing narrow, high-velocity emission-line components at the redshift of \sourceshortname\ \citep[e.g.][]{schneider_1985_paperii,Schneider1986}.~\cite{Yamada2001} subsequently analyzed \cfht\ spectra of images B and C, detecting \lyalpha, \nitrogenfive, \carbonfour, \heliumtwo, and \carbonthree\ in both images, while additionally finding significant emission-line ratio differences between images B and C (notably, in \heliumtwo/\carbonfour\ and \carbonthree/\carbonfour).~Adopting the single-source lens models for \sourceshortname\ available at the time, these emission-line ratio differences were interpreted as arising from variations in the ionization parameter across a narrow-line region of a single type-II lensed AGN \citep{Yamada2001}.~While the spatially distinct radio sub-components of images A--C were initially associated with radio jets emanating from a single AGN \citep[][]{Koopmans2002b}, the complex radio morphology of image C (shown in Figure \ref{fig:figure1_hst+muse}) became a matter of debate.~The first global VLBI observations of \sourceshortname\ were used to extend previous single-source lens models by integrating both the main lensing galaxy and a satellite galaxy into the effective lensing gravitational potential \citep{more2009}.

Later, the multi-epoch VLBI study of \citet{Spingola2019} suggested that \sourceshortname\ comprises two sources, where each ``source’’ denotes a compact radio core and its associated jet, separated by a projected distance of \offset\ at the source redshift.~The source-plane reconstruction and spectral interpretation of \citet{Spingola2019}~strongly favored an unusual two-source lensing geometry:~``source 1’’ lies within the inner lensing caustic and is quadruply lensed and highly magnified ($\mu_\mathrm{C}\sim350$), where two of its lensed images are in the merging configuration, producing C, while ``source 2’’ lies outside the caustic and produces A and B ($\mu_\mathrm{A,B}\sim2$).~Although \citet{Spingola2019} could not rule out a single-AGN interpretation in which C arises from a complex jet structure from source 2 crossing the inner caustic, the astrometric X-ray-to-radio determination of \citet{Schwartz2021} subsequently identified two spatially distinct X-ray sources coincident with these two radio cores.~New \axaf\ observations reveal strong intrinsic X-ray absorption toward C ($N_\mathrm{H}>4\times10^{22}\ \mathrm{cm}^{-2}$ at the $3\sigma$ level) at $z=$~\sourceredshift~\citep{schwartz_mgb2016_2026}. This level of absorption cannot be reconciled with reprocessing from obscuring material surrounding an AGN \offset\ away from C (in projection), providing further evidence that source 1 is an obscured AGN in addition to the AGN associated with A and B~\citep[i.e.~source 2; ][]{schwartz_mgb2016_2026}.

In this work, we exploit \muse\ observations of \sourceshortname\ to present the first simultaneous, spatially resolved spectroscopic analysis of its distinct lensed images (A--C).~We investigate their emission-line properties -- including the effects of intervening and proximate absorption -- and perform spectral fitting of their  \lyalpha\ and \carbonfour\ emission regions.~Using the \carbonfour\ emission line, we place the first constraints on the SMBH masses associated with each lensed image.~Together with existing predictions for the strong lensing magnifications,~combined with the rest-frame luminosities inferred from the \muse\ data and recent \axaf\ observations,~we use our SMBH mass lower limits to estimate their Eddington ratios.~Our investigation provides new insights into the environmental and physical properties of the two SMBHs in this unique lensed dual AGN.

Our analysis provides strong, independent spectroscopic evidence for two distinct ionizing sources in \sourcefullname\ from the statistically significant differences in the \heliumtwo/\carbonthree\ emission-line ratios for images A/B vs. C.~{Our findings}~motivate rest-frame optical follow-up to obtain independent SMBH mass estimates from other emission lines that are unaffected by saturated absorption, and to characterize the host galaxy environment.~The \muse\ data analyzed here provide insight into the complex emission and absorption in \sourceshortname~--~a compact dual AGN system that is uniquely revealed by strong gravitational lensing.

\section{Discovery and overview of \sourcefullname}
\label{sec:sec2_intro-to-systems}

\begin{figure*}
    \centering
    \begin{subfigure}[t]{0.58\linewidth}
        \centering
        \includegraphics[width=.805\linewidth,height=0.65\linewidth]{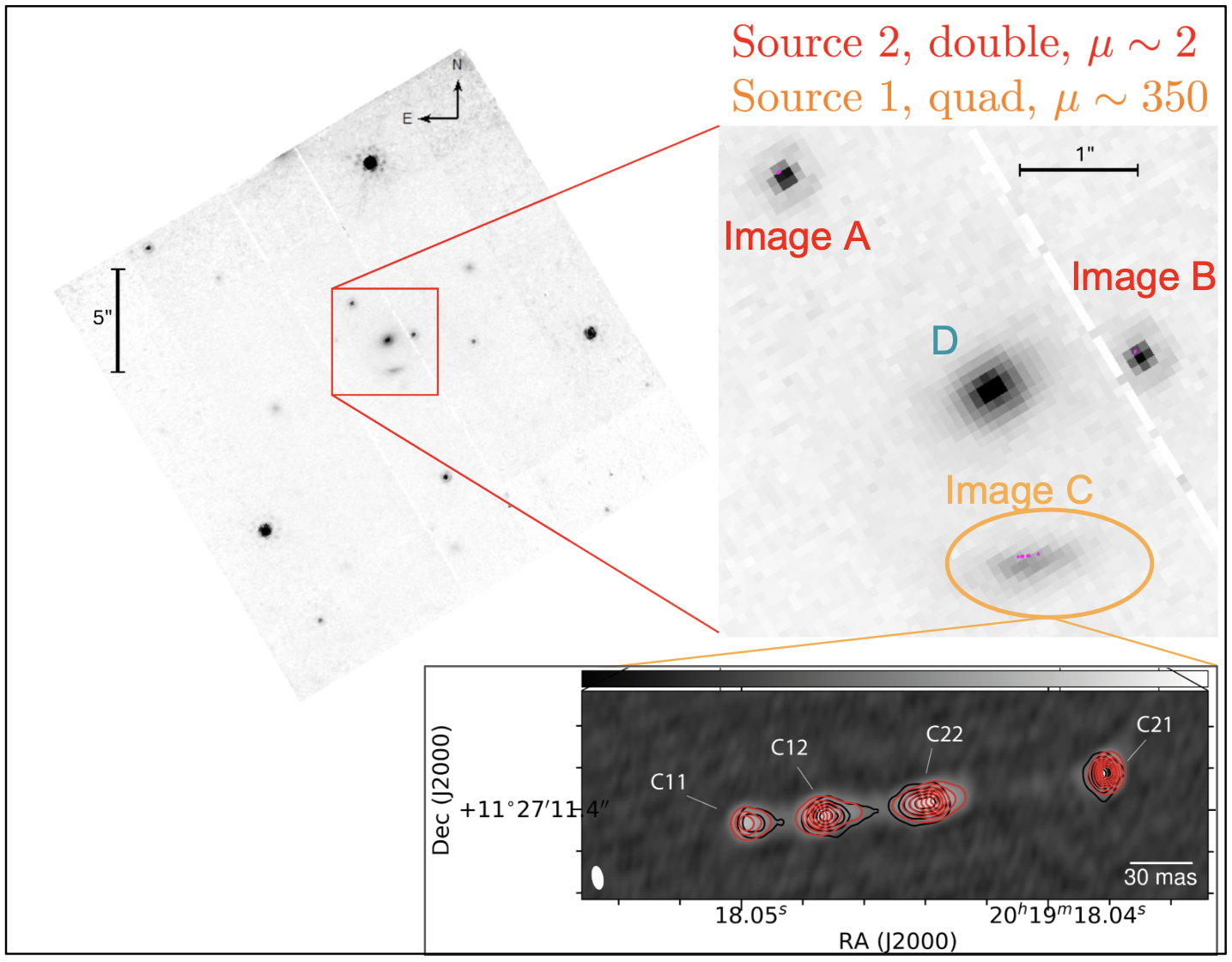}
        \label{fig:hst_fov}
    \end{subfigure}
    \hfill
    \begin{subfigure}[t]{0.34\linewidth}
        \centering
        \setlength{\fboxsep}{0pt}
        \fbox{%
            \includegraphics[width=1\linewidth,height=1.1\linewidth]{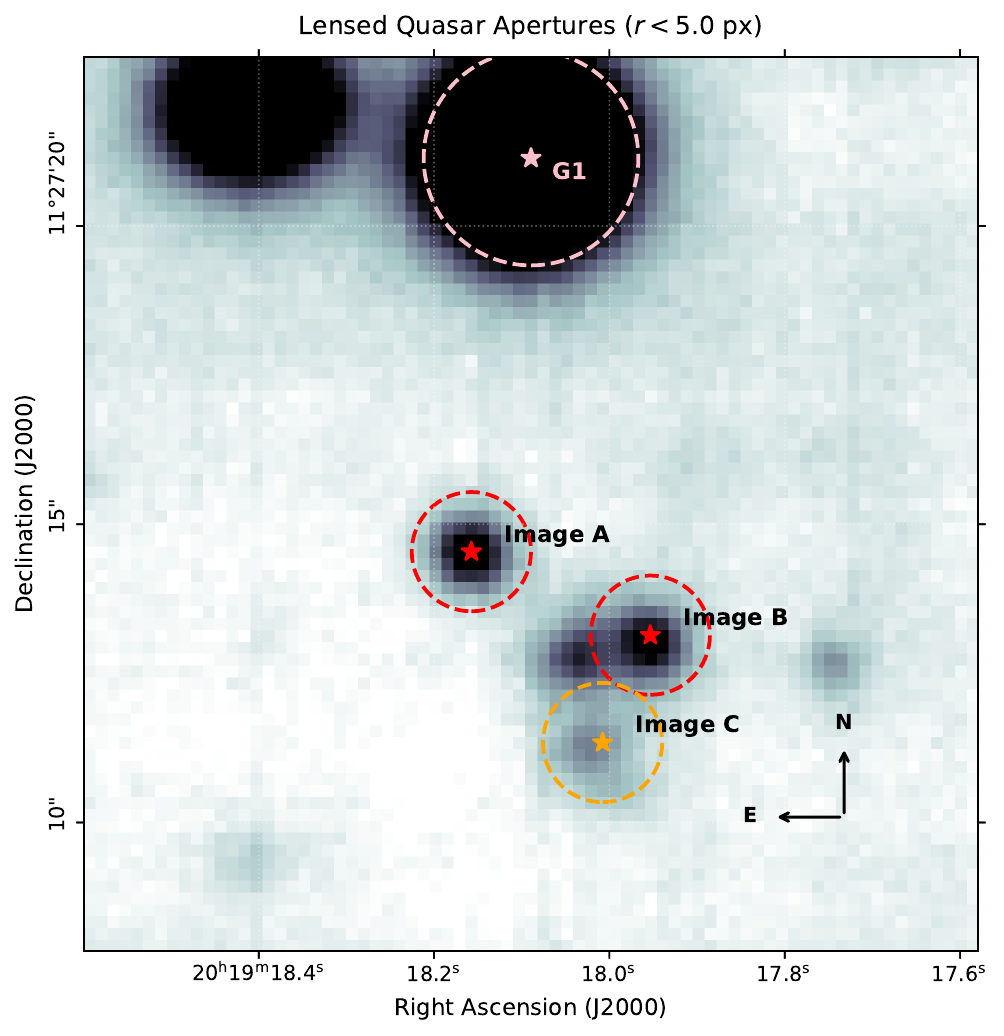}%
        }
        \label{fig:muse_fov}
    \end{subfigure}

    \vspace*{-0.2cm}
    \caption{Composite \hst/NICMOS (left) and \muse/X-IFU (right) FOV views of \sourcefullname.~Both panels are oriented following the standard convention where up corresponds to North, and left to East.~Both panels are centered approximately at J2000 RA 29h 19m 18.25s and Dec +11$^{\circ}$ 27' 17.04"~(the light centroid of the lensing galaxy),~and represent $\approx 5\farcs0\times 5\farcs0$ and $\approx 20\farcs0\times 20\farcs0$ cutouts, respectively.~The left panel marks the image-plane configuration:~the three optical lensed images (A--C) and the main lensing galaxy D are spatially resolved by \muse.~The magenta contours on image C are taken from the 1.7 GHz VLBA image presented by \citet[][]{Spingola2019}, accessed via Vizier~(DOI: \href{10.26093/cds/vizier.36300108}{10.26093/cds/vizier.36300108}).~The inset into image C shows its sub-components (C11-C12, and C21-C22), and is taken directly from Figure 1 of \cite{Spingola2019}.~The legend indicates the dual AGN lens model favored by \citet{Spingola2019}:~images A and B arise from source 2 (doubly lensed), whereas C arises from source 1 (quadruply lensed).~The \muse\ FOV (right) shows the galaxy G1, whose spectra we used to correct for $z\sim 0$ absorption prior to extracting emission-line spectra for images A--C~(see Section \ref{sec:sec3_extraction}).}
    \label{fig:figure1_hst+muse}
\end{figure*}

\sourcefullname\ was serendipitously discovered by the MIT Green Bank Survey as the fourth confirmed gravitational lens \citep{Lawrence1984}~and has been the subject of extensive lens modeling efforts~\citep{narashima_1984_glmodel,Schneider1986,lawrence1993,garrett_1994_merlin,Yamada2001,Koopmans2002a,Koopmans2002b,more2009,Spingola2019}.~\sourceshortname\ has historically been classified as a type-II system at optical,~radio,~and X-ray wavelengths \citep[e.g.][]{Yamada2001,Koopmans2002b,schwartz_mgb2016_2026}.

Figure \ref{fig:figure1_hst+muse} shows the \hst/NICMOS configuration of \sourceshortname\ obtained in July 1997 with an effective 85-minute exposure (PI:~E.~Falco; ObsID 7495; filters:~F160W; DOI:~\href{https://archive.stsci.edu/doi/resolve/resolve.html?doi=10.17909/mqwz-2q10}
{10.17909/mqwz-2q10}).~The three macroscopic lensed images (A--C) are well-resolved at a maximum separation of \maxsephst.~The lensing galaxy, D, is a massive elliptical at $z = 1.001$ with an old, metal-rich stellar population and a measured stellar velocity dispersion of $\sim 330 \pm 30 \ \mathrm{km~s^{-1}}$ \citep{Koopmans2002a}.~An optical spectroscopic survey by \citet{soucail_2001} identified an overdensity of six galaxies at $z\sim 1$, suggesting that D resides in a larger-scale structure (e.g.~a galaxy group or cluster).~While the discovery of an X-ray source at the position of the overdensity by the \asca\ satellite (with $\sim1$ arcmin angular resolution) initially favored a cluster interpretation \citep[][]{Hattori97}, the first \axaf\ observation of \sourceshortname\ (with $\sim 8 \ \mathrm{ks}$ of exposure)~clearly resolved three point-like images and found no significant diffuse X-ray emission associated with the overdensity, therefore disfavoring the presence of a massive, virialized cluster \citep{Chartas2001}.

Several lens models have been proposed for \sourceshortname\ to explain its unusual radio morphology, as summarized in Section \ref{sec:sec1_intro} \citep[and also reviewed extensively in section 2 of][]{more2009}.~Recently,~\citet{Spingola2019} used multi-epoch VLBI observations \citep[including those used in the first global VLBI study of][]{more2009}~to reconstruct the radio emitting regions in the source plane.~After identifying four distinct radio components in the source plane, \citet{Spingola2019} considered several possible lensed single AGN and dual AGN configurations, strongly favoring a lens configuration where the source is a dual AGN (where each AGN has its own radio jet) at a projected separation of \offset.~This two-source scenario is independently supported by our \muse\ spectral analysis.

The archival \muse\ observations analyzed here provide a substantial improvement from the early spectroscopic \cfht\ study of images B and C by \cite{Yamada2001}.~Compared with these previous observations on \cfht\ with the Subarcsecond
Imaging Spectrograph (SIS) and R150 grating, our extracted \muse\ spectra achieve a spectral resolution $\gtrsim5$ times higher than  \cfht\ ($\sim 2\,989$ for VLT/\muse\ vs.~$\sim400{-}600$ for \cfht/SIS).~The improved spectral resolution of \muse\ allows us to put better constraints on the emission-line widths and, for the first time, to place lower limits on the individual SMBH masses associated with images A--C.~We note that the \cfht\ analysis of \cite{Yamada2001} had already reported the detection of \lyalpha, \nitrogenfive, \carbonfour, \heliumtwo, and \carbonthree\ across images B and C, all of which were fitted with a single, narrow Gaussian component.~Previously,~\cite{schneider_1985_paperii}~had used the Hale optical telescope and detected~\lyalpha,~\nitrogenfive\,~\carbonfour,~and \heliumtwo~in images A and B.

The remaining sections of this paper are organized as follows.~Section \ref{sec:sec3_extraction}~outlines our data extraction.~Section \ref{sec:sec4_results+discussion}~presents our spectroscopic analyses of images A--C.~We present the detection of the various emission lines and determine their corresponding emission-line ratios (Section \ref{subsec:sec_emission-lines}).~We then perform multi-Gaussian fitting of the \lyalpha\ and \carbonfour\ emission lines.~Using the broad component of \carbonfour, we subsequently place lower limits on the black hole masses associated with the three lensed images (Section \ref{subsubsec:sec_smbh-masses}).~Finally, we combine these SMBH mass estimates along with the strong lensing magnifications $\mu$ predicted by the two-source lens model of~\citet{Spingola2019}~and their observed 1\,350\AA\ luminosities and intrinsic, unabsorbed X-ray luminosities to estimate their respective Eddington ratios (Section \ref{subsubsec:sec_edd-ratios}). We summarize our findings and conclude in Section \ref{sec:sec5_conclusions}.

\begin{figure}
\includegraphics[height=6cm,width=18cm]{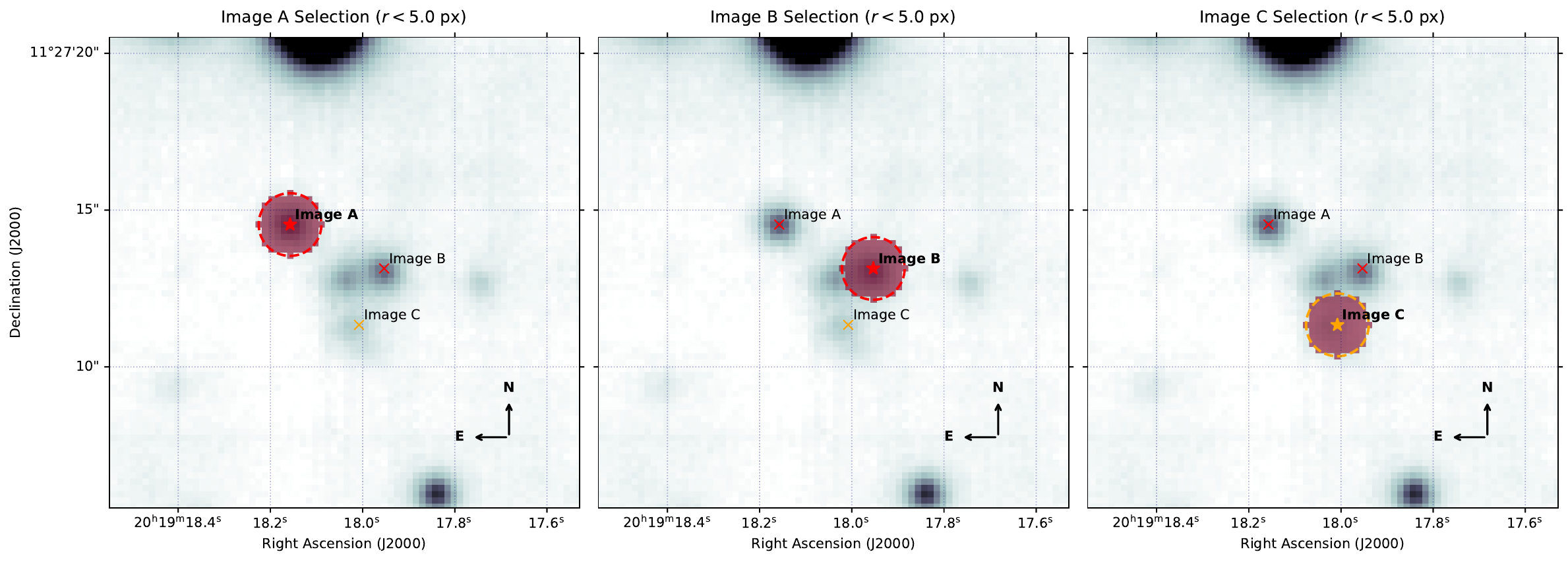}
\caption{\muse/X-IFU FOV views of \sourcefullname, showing the 5-pixel apertures in radius we integrated across to obtain the individual spectra of images A (left), B (center), and C (right).~The three panels are oriented following the standard convention where up is North and left is East.~The three panels are centered approximately at J2000 RA 29h 19m 18.25s and Dec +11$^{\circ}$ 27' 17.04"~(the light centroid of the lensing galaxy), and represent $\approx 15\farcs0\times 15\farcs0$ cutouts.~The lensing galaxy (D in Figure \ref{fig:figure1_hst+muse}) is well-resolved and the North-most object is G1 (see Section \ref{sec:sec3_extraction} for details).}
\label{fig:figure2_muse}
\end{figure}

\section{Data extraction and analysis}
\label{sec:sec3_extraction}

The accessible 3-dimensional \muse\ data cube analyzed here is part of an observing run that took place on August 5, 2018 \citep[PID:~2101.B-5037(A); PI:~C.~Spingola;][]{ESO_2017}.~These data have the following specifications:~total exposure time of $5\,575\;\mathrm{s}$; covering field of view (FOV) $\sim 1' \times 1'$; angular FWHM:~$1\farcs09$; wavelength range:~$\sim 4\,600\mathrm{\AA}$ to $9\,350\mathrm{\AA}$.~We extracted the spectra of the individual lensed images (A--C) by integrating the flux within a 5-pixel circular aperture in radius centered on the peak flux pixel in the VLT/\muse~data cube ($0\farcs2$/pixel; Figure \ref{fig:figure2_muse}).~This aperture was chosen to boost the S/N for image C and to calculate the integrated 1\,350 \AA\ luminosity for images A--C.~While we do not expect the lensing galaxy D (a massive elliptical) to affect the emission-line spectra we extracted for images A--C (as described below), D could potentially affect the continuum absorption.

The galaxies around \sourceshortname\ in the \muse\ FOV are found to be absorbed by both the local interstellar medium (ISM; i.e.~at $z\sim0$) and a set of proximate absorbers to \sourceshortname~($z\sim$~3.27), which we detail below.~To obtain the individual emission-line spectra for images A--C, we thus corrected for the absorption components at $z\sim0$, and masked the absorption lines for those at $z\sim$~\ztproximateabs.~While the lensing galaxy D may also have absorption features that are additionally blending with the spectra of images A--C, we do not find significant absorption lines at $z=1.001$ in their spectra \citep[as expected, since D is a passive elliptical galaxy;~see e.g.][]{soucail_2001}.

The \magnesiumone\ $\lambda \lambda \lambda \ 5168,~5174,~5185$~absorption lines at $z\sim 0$ are present in the spectra of multiple galaxies and objects across the \muse\ field of view (FOV), where they overlap with and attenuate the \lyalpha\ emission line of images A--C~(observed at $\sim 5150 – 5120$ \AA;~Figure \ref{fig:spectral-fitting}).~Correcting for this foreground absorption is therefore necessary to characterize the intrinsic \lyalpha\ emission-line fluxes and profiles across images A--C. To account for this effect, we selected G1 (Figure~\ref{fig:figure2_muse})~--~a galaxy with the smallest angular separation from the lensed images and located approximately $\sim 8\farcs0$~from image A~--~as an empirical reference for the local ISM absorption.~NED does not report a redshift for G1,~and its \muse\ spectrum does not reveal identifiable emission lines from which its redshift can be determined.~Since it is very unlikely that G1 is associated with the dual AGN -- given its angular separation to \sourceshortname\ and the absence of emission lines in its spectrum --, we use it as a reference galaxy to correct for \magnesiumone\ absorption of the local ISM at $z\sim 0$. The physical properties of G1 -- whose determination is beyond the scope of this paper -- do not affect the physical interpretation of the dual AGN system \sourceshortname\ presented here.

We also identify the presence of proximate absorbers along the line of sight to the quasar~-- specifically,~a set of \lyalpha, \nitrogenfive, and \carbonfour\ absorption lines at $z\approx$~\sourceredshift.~To robustly characterize the emission lines of \sourceshortname\ across images A--C, we masked the spectral regions affected by these absorption features when fitting the \lyalpha, \nitrogenfive,~and \carbonfour\ emission-line profiles in the individual image spectra.~We note that these masks were not applied to estimate the emission-line fluxes across images A--C (subsequently used to compute their respective emission-line ratios; Table \ref{tab:line_ratio_comparison}).

\section{Results and discussion}
\label{sec:sec4_results+discussion}

\subsection{Emission-line detections and emission-line ratios for \sourcefullname}
\label{subsec:sec_emission-lines}

\begin{figure}
\includegraphics[width=\textwidth]{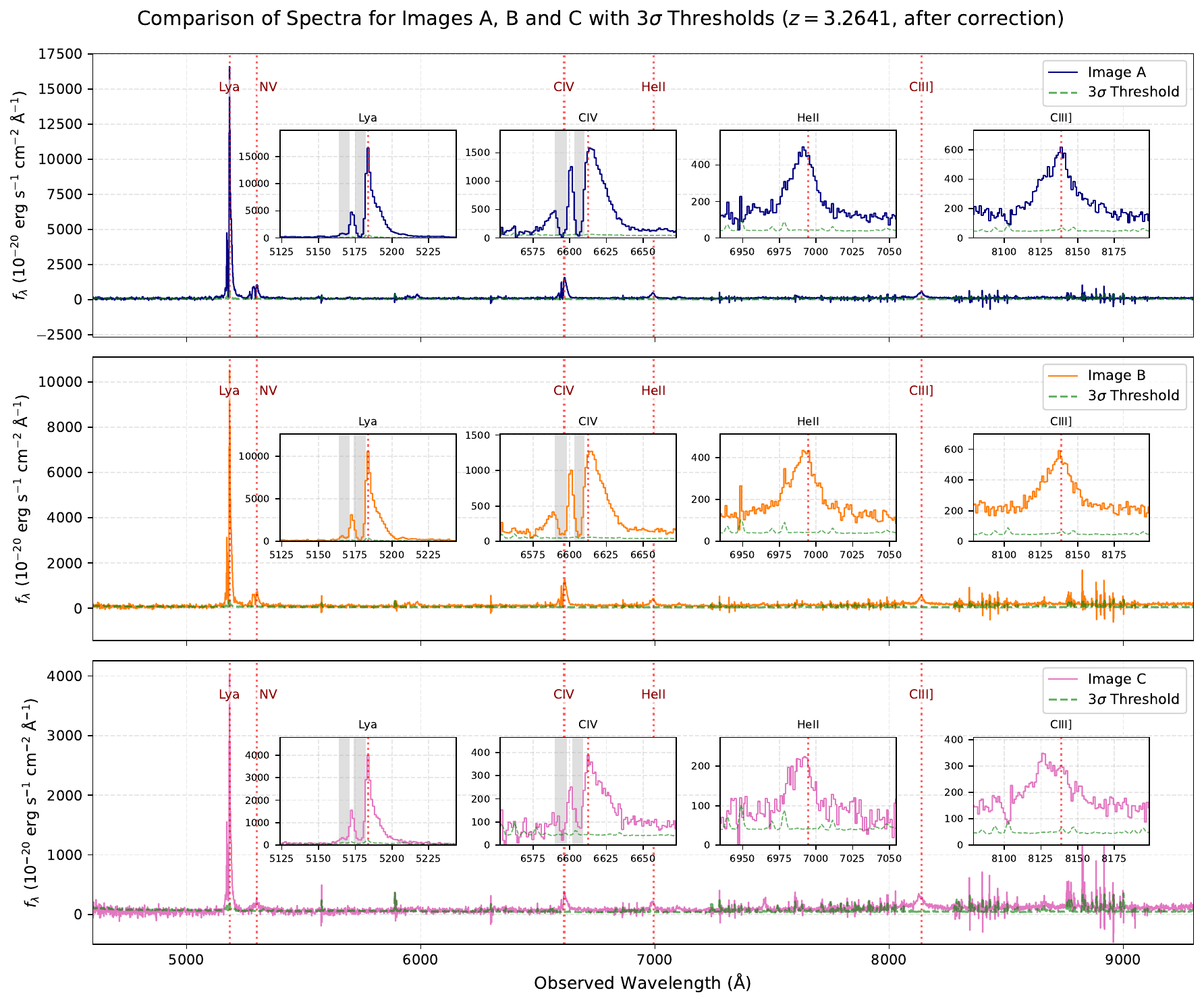}
\caption{
Wavelength-aligned \muse\ spectra of lensed images A--C (navy:~image A, top; orange:~image B, center; pink:~image C, bottom).~The $3\sigma$ flux thresholds in all three panels are plotted in dashed green lines.~The red vertical dotted lines across images A--C show the emission lines detected at a rest-frame redshift $z=3.2641$.~\lyalpha, \nitrogenfive, \carbonfour, \heliumtwo\ and \carbonthree\ emission lines are all detected in images A, B, and C.
The inset panels display zoom-in views of the \lyalpha,~\carbonfour,~\heliumtwo\ and \carbonthree\ emission-line regions.~The shaded grey areas in the \lyalpha\ and \carbonfour~insets mark regions affected by the proximate absorbers (described in Section~\ref{sec:sec3_extraction}).
} 
\label{fig:figure3_spec}
\end{figure}

We identify multiple emission lines -- including \lyalpha, \nitrogenfive, \carbonfour, \heliumtwo, and \carbonthree\ -- across images A--C, as shown in Figure \ref{fig:figure3_spec}.~These emission lines had previously been detected in \cfht\ observations of images B and C by \citet[][]{Yamada2001}.~Figure \ref{fig:figure3_spec} also illustrates that the spectrum of image C is considerably noisier than those of A and B. 

We find that the absorption features associated with \lyalpha, \nitrogenfive, and \carbonfour\ at a redshift of $z\sim$~\ztproximateabs\ are saturated (as described in Section \ref{sec:sec3_extraction}).~The combination of blending and saturation prevents a robust determination of their column densities.

Table \ref{tab:line_ratio_comparison} presents the emission-line flux ratios measured from the \muse\ spectra of images A--C after correcting for absorption at $z\sim 0$ (as described in Section~\ref{sec:sec3_extraction}).~The \lyalpha/\carbonfour\ ratio for images A and B and the \carbonfour/\carbonthree\ ratio for images A/B and C display notable differences. While discrepancies in the latter ratio for images B and C had already been reported by \citet{Yamada2001}, here both \lyalpha\ and \carbonfour\ are affected by saturated absorption, which can introduce uncertainties and potentially image-dependent differences in their emission-line ratios.~Therefore, we defer a detailed investigation of these discrepancies to future work and instead focus on thee \heliumtwo/\carbonthree\ ratio -- the only ratio computed here that is unaffected by saturated absorption and therefore provides a more robust comparison of the intrinsic emission-line properties. 

The \heliumtwo/\carbonthree\ emission-line ratio is consistent between images A and B, as expected if they represent two lensed images of the same background source, but differs significantly between images A/B and C.~A simple $\chi^2$ test under the null hypothesis that the measured \heliumtwo/\carbonthree\ ratios are constant across the three images yields $\chi^2 = 110$ for 3 degrees of freedom, strongly rejecting this hypothesis.~This simple $\chi^2$ test provides strong additional spectroscopic evidence that image C is associated with an ionizing source distinct from that producing images A and B -- expected given the dual AGN nature of \sourceshortname.~While \citet{Yamada2001}~noted that spatially extended, ionization-stratified narrow-line region emission may contribute to differences in the emission-line ratios they measured for images B and C,~the \muse\ data cannot directly quantitatively assess this scenario.

\subsection{Physical properties of the accreting system in \sourcefullname\ from emission-line properties}
\label{subsec:sec_emission-lines_physical-properties}

\begin{figure}[htbp]
    \centering
    \includegraphics[width=\textwidth]{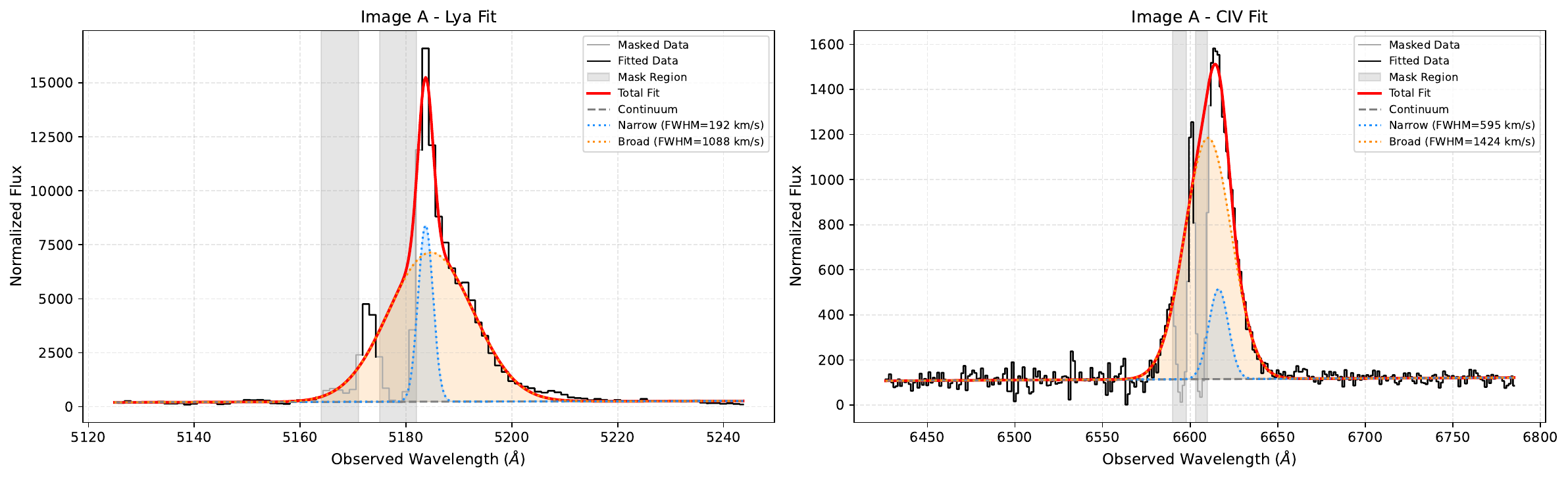} \\[1ex]
    \includegraphics[width=\textwidth]{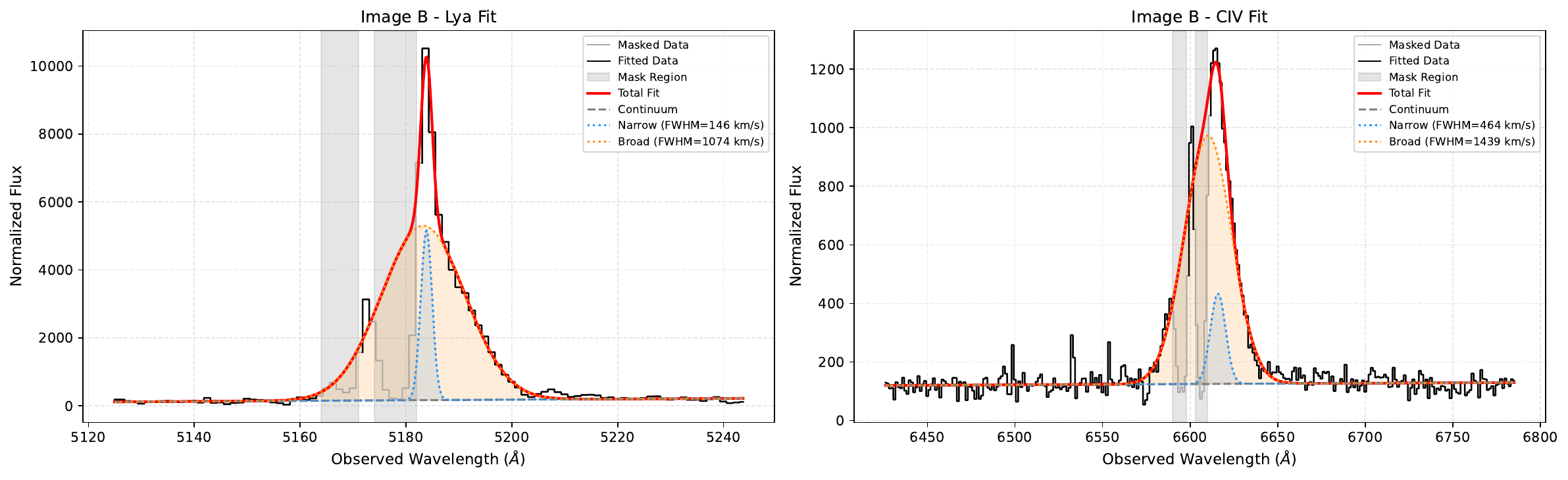} \\[1ex]
    \includegraphics[width=\textwidth]{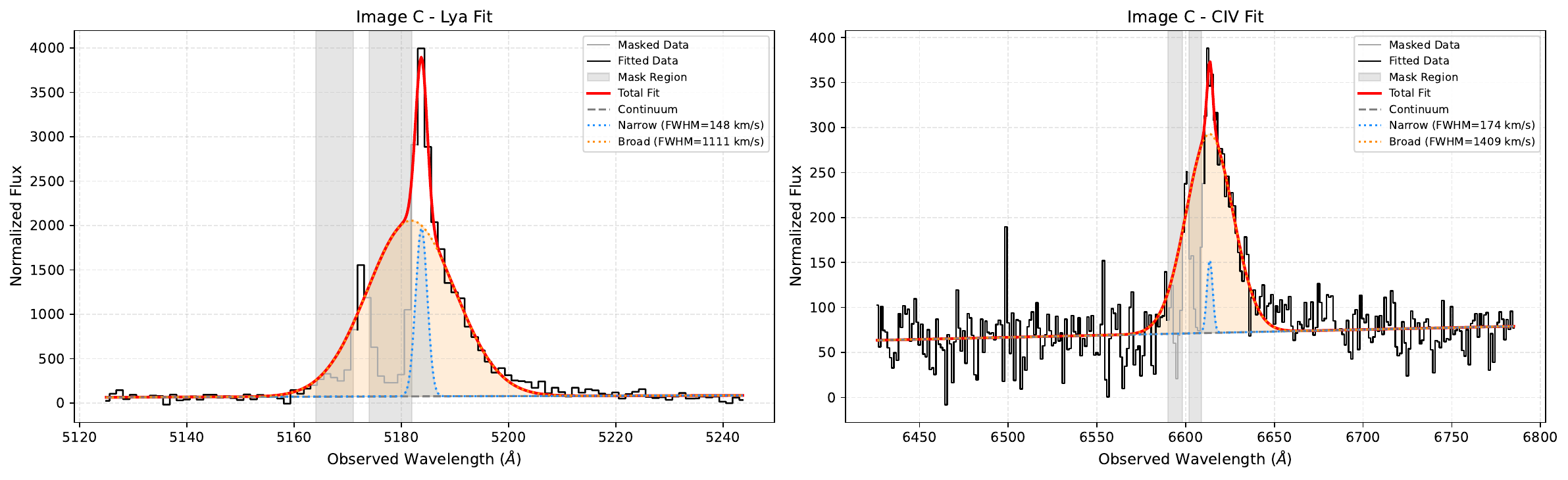}
    \caption{Gaussian fits of the \lyalpha\ and \carbonfour\ emission lines from \muse\ across in images A (top), B (center), and C (bottom).~All Gaussian fits were multi-component, requiring both a narrow (blue) and broad (orange) components.~The best-fit parameters are summarized in Table \ref{tab:fitted_parameters}.~Table \ref{tab:fitted_parameters} also shows the lower SMBH mass limits estimated from the broad \carbonfour\ component in images A--C.~The regions of absorption due to proximate absorbers (marked by the gray shaded regions in all panels) are masked during the fitting, as described in Sec.~\ref{subsec:sec_emission-lines}.
    }
    \label{fig:spectral-fitting}
\end{figure}

The high-quality \muse\ spectroscopy analyzed here provides a unique opportunity to characterize the emission-line properties of the individual lensed images A--C.~As shown in Figure~\ref{fig:spectral-fitting},~we model the \lyalpha\ and \carbonfour\ emission-line regions across images A--C using a combination of a linear continuum and multiple Gaussian components to account for the local continuum and line emission, respectively.~The optimal number of components is determined through visual inspection, guided by physical priors on the expected line kinematics and morphology.~Following common practice in IFU spectral fitting, we allow for two Gaussian components where warranted by the data, particularly when distinct narrow and broad emission-line components are revealed, as is the case for images A--C (Figures~\ref{fig:figure3_spec} and \ref{fig:spectral-fitting}).~We also tested a more complex model comprising one narrow and two broad Gaussian components.~However, the additional broad component does not significantly improve the emission-line profile descriptions and only changes the inferred SMBH masses (presented in Section~\ref{subsubsec:sec_smbh-masses})~by $0.1 \ \mathrm{dex}$.~This additional component primarily accommodates uncertainties in the local continuum or in the spectral regions masked due to absorption by the proximate absorbers at $z\sim3.273$,~which are not sufficiently constrained to justify the additional model complexity.~We therefore adopt a model comprising one narrow and one broad Gaussian component for the \lyalpha\ and \carbonfour\ emission regions in images A--C.

Model parameters including amplitudes,~central wavelengths,~and velocity dispersions ($\sigma$) are optimized with least-squares minimization using the default Trust Region Reflective (TRF) algorithm implemented in \texttt{scipy.optimize.curve{\_}fit}.~The wavelength masks are also applied (shown as grey shades in Figure~\ref{fig:spectral-fitting}) to exclude the absorption-line regions due to the proximate absorbers at $z\sim3.273$.~Table~\ref{tab:fitted_parameters}~reports the best-fit properties of the narrow and broad components of the \lyalpha\ and \carbonfour\ emission regions across images A--C, including the effective velocity dispersions ($\sigma$) of the broad components.~The FWHM of each component is calculated as $\mathrm{FWHM}=2.35\sigma$.~Table \ref{tab:fitted_parameters}~also shows the monochromatic continuum fluxes of images A--C (see also Figure \ref{fig:figure3_spec}).

Although the moderately broad Gaussian components in the \carbonfour\ emission-line fits ($\sigma \sim 580 - 618 \ \mathrm{km~{s}^{-1}}$; Table~\ref{tab:fitted_parameters})~do not, by themselves, establish a BLR origin, as already noted by \citet{Yamada2001} -- who argued \sourceshortname\ is a type-II quasar --, the AGN-like nature of images A--C motivates us to explore a virial interpretation of their line widths.~Furthermore, even if the velocity dispersions could be confidently attributed to the BLR (i.e., if $\sigma > 3\,000 \ \mathrm{km~{s}^{-1}}$), the resulting virial SMBH mass estimates based on \carbonfour\ would still need to be carefully assessed. For example, single-epoch virial black hole mass estimates based on broadened \carbonfour\ lines are known to be overestimated when the lines exhibit strong blueshifts \citep{coatman_2017_i,coatman_2017_ii}.~We therefore adopt deliberately large uncertainties to account for the additional scatter in \carbonfour\ virial mass estimates relative to other SMBH mass proxies (Table \ref{tab:fitted_parameters}).

We use the velocity dispersions of the broad Gaussian \carbonfour\ components in virial scaling relations calibrated for the BLR in type-I AGN to derive lower limits on the SMBH masses associated with the individual lensed images~(Section~\ref{subsubsec:sec_smbh-masses}).~To the best of our knowledge, this represents the first attempt towards estimating the individual masses of the two SMBHs in this compact dual AGN system.~Combining these mass estimates with the monochromatic continuum fluxes at 1\,350~\AA~of images A--C -- corrected with the average magnification factors predicted by the two-source model of \citet{Spingola2019} -- further allows us to estimate their respective Eddington ratios (Section \ref{subsubsec:sec_edd-ratios}).~These limits are consistent with those independently derived from the intrinsic X-ray luminosities of sources 1 and 2 \citep[which we estimate from][]{schwartz_mgb2016_2026},~assuming a reasonable X-ray bolometric correction and adopting our SMBH mass lower limits (Section \ref{subsubsec:sec_edd-ratios}).~Both our SMBH mass and Eddington ratio estimates suggest additional constraints on the phenomenological properties of the dual AGN system and its accretion state.

\subsubsection{Black hole masses}
\label{subsubsec:sec_smbh-masses}
We estimate the black hole masses using the single-epoch \carbonfour\ virial scaling relation of \citet{2006ApJ...641..689V}, using the observed~\carbonfour\ broad-line widths and the integrated (observed) 1\,350\ \AA\ luminosities found for images A -- C:~$4.3\times 10^{44} \ \mathrm{erg~{s}^{-1}}$, $5.3\times 10^{44} \ \mathrm{erg~{s}^{-1}}$, and $3.5\times 10^{44} \ \mathrm{erg~{s}^{-1}}$,~respectively.~This relation,~calibrated against SMBH masses inferred from reverberation mapping, follows:

\begin{equation} 
\label{eq:eq0}
\mathrm{log}\{M_\mathrm{BH}(\carbonfour)\} = \mathrm{log} \Big\{ [\frac{\mathrm{FWHM}~(\carbonfour)}{1\, 000 \ {\rm km/s}}]^{2}~ [\frac{\lambda L_{\lambda}(1\,350\ \mathring{\rm A})}{10^{44} \ {\rm erg/s}}]^{0.53} \Big\} + (6.66\pm 0.01),
\end{equation}

\noindent and exhibits a scatter of $\approx 0.36 \ \mathrm{dex}$ relative to reverberation-mapping-based masses.~Additionally accounting for the uncertainties in the reverberation mapping mass scale, an absolute statistical uncertainty of $0.56 \ \mathrm{dex}$ for the \carbonfour\ FWHM-based mass estimator is reported in table 5 of \citet[][]{2006ApJ...641..689V}.

The scatter associated with comparing the single-epoch virial scaling relation in Equation \ref{eq:eq1}~with other mass proxies dominates the uncertainty budget over the formal uncertainties in the FWHM and 1\,350\AA\ luminosities from our emission-line fits. We therefore adopt an uncertainty of $\approx 0.56 \ \mathrm{dex}$ for the SMBH masses reported in Table \ref{tab:fitted_parameters}, which already account for this intrinsic scatter.

Since the scaling relation in Equation \ref{eq:eq0}~is calibrated for unobscured, type-I AGN, its application to the broad component detected in an otherwise apparently narrow-line (type-II) source may introduce additional systematic uncertainties that are not captured by the adopted scatter.~That is,~if the observed \carbonfour\ emission were indeed originating from outside the BLR and traced lower velocities than the intrinsic virialized BLR gas, applying the broad-line virial scaling relation in Equation~\ref{eq:eq0}~would underestimate the underlying SMBH masses.~Moreover,~the integrated (observed) 1\,350\AA\ luminosities used in Equation \ref{eq:eq0} are underestimated,~as these do not represent the absorption-corrected values.~We were not able to compute the latter because of the strong, saturated absorption present in \lyalpha.~Therefore, the SMBH estimates may be regarded as approximate lower limits in such scenario.~In parallel, because the observed \carbonfour\ line widths may also be influenced by non-virial gas motions, the SMBH mass estimates remain model-dependent.

Table \ref{tab:fitted_parameters} shows the lower-limit SMBH mass estimates for images A--C:~$M_\mathrm{SMBH,A} \gtrsim$~\massalowlim,~$M_\mathrm{SMBH,B} \gtrsim$~\massblowlim,~and $M_\mathrm{SMBH,C} \gtrsim$~\massclowlim.~If these lower limits are representative of the true underlying SMBH masses,~they would suggest that the dual SMBH in \sourceshortname\ has a relatively high mass ratio,~potentially indicating that the host galaxy has undergone a major merger.~However,~this interpretation cannot be robustly investigated with the available \muse\ data,~and would first require reliable SMBH mass measurements for each lensed image.

At a projected separation of only \offset\ at $z=$~\sourceredshift, similar-separation dual AGN states remain in a parameter space that is largely unexplored by cosmological models of structure formation \citep[e.g.][]{chen_2023_astrid}.~This makes \sourceshortname\ a unique laboratory for investigating the properties and evolution of close-separation SMBH pairs at high $z$ and motivates deeper spectroscopic follow-up to further characterize its SMBH masses,~to potentially identify tracers of star formation in the host galaxy,~and to assess its broader theoretical implications.

\begin{table}[htbp]
    \centering
    \caption{Comparison of the emission-flux ratios for images A, B, and C measured from the \muse\ spectra.~Their $1\sigma$ statistical uncertainties were estimated via standard error propagation of the statistical uncertainties on the observed emission-line fluxes.~No uncertainties are reported for the emission-line ratios, as the dominant source of uncertainty for all individual emission-line regions is the \muse\ spectral resolution ($\approx 2.5$–$2.7$ \AA).}
    
    \label{tab:line_ratio_comparison}
    \begin{tabular}{lcccc}
        \hline\hline
        Spectrum & \lyalpha/\textnormal{C\,{\sc iv}} & \carbonfour/\heliumtwo & \carbonfour/\carbonthree & \heliumtwo/\carbonthree \\
        \hline
        Image A & $3.7$ & $3.3$ & $2.0$ & $0.6$ \\
        Image B & $2.7$ & $3.1$ & $1.8$ & $0.6$ \\
        Image C & $3.7$ & $2.0$ & $0.7$ & $0.4$ \\
        \hline
    \end{tabular}
\end{table}

\begin{table*}[htbp]
    \centering
    \caption{Summary of our two-Gaussian (narrow+broad) fits to the \lyalpha\ and \carbonfour\ emission lines in images A--C.~The FWHM of all Gaussians is $\mathrm{FWHM} = 2.35\sigma$.~In cases where uncertainties for the fitted parameters are reported,~these are calculated as described in Section~\ref{subsec:sec_emission-lines_physical-properties}.~In cases where no uncertainty is presented,~the underlying uncertainty is dominated by the \muse\ spectral resolution ($\approx2.5$--$2.7$~\AA).~The last column reports the corresponding SMBH mass lower limits.~As detailed in Section \ref{subsubsec:sec_smbh-masses},~the~$0.56 \ \mathrm{dex}$~uncertainty assumed in $\mathrm{log}(M_\mathrm{BH}/M_\odot)$~--~shown in the second-to-last-column~-- corresponds to the scatter in the \carbonfour~single-epoch virial scaling relation of \citet{2006ApJ...641..689V}~found when comparing this relation to other SMBH mass scaling relations.~The observed 1\,350 \AA\ continuum flux densities of images A--C (integrated through a 5-pixel aperture, and determined by the mean flux of 20 pixels at rest-frame 1\,350\AA)~are also listed.~Section \ref{subsubsec:sec_smbh-masses}~lists the integrated (observed) 1\,350\AA\ luminosities for images A--C inferred from these continuum flux densities.}
    \label{tab:fitted_parameters}
    \small
    \resizebox{\textwidth}{!}{%
    \begin{tabular}{llcccccc}
        \hline\hline
        Image & Line
        & Narrow Center & Narrow $\sigma$
        & Broad Center & Broad $\sigma$
        & $f_{1350}$
        & $M_{\mathrm{BH}}$ \\
        & & (\AA) & (km\,s$^{-1}$)
        & (\AA) & (km\,s$^{-1}$)
        & ($10^{-20}$ erg\,s$^{-1}$\,cm$^{-2}$\,\AA$^{-1}$) & ($M_\odot$) \\
        \hline
        Image A & Ly$\alpha$
        & $5\,184$ & $82 \pm 2$
        & $5\,185$ & $462 \pm 2$
        & -- & -- \\
        & C\,IV
        & $6\,616$ & $253 \pm 14$
        & $6\,610$ & $605 \pm 6$
        &  -- & $\gtrsim 5.5 \times 10^6$ \\
          & Continuum & -- & --  & -- & -- & $76\pm 4$ &--\\[2pt]
         Image B & Ly$\alpha$
        & $5\,184$ & $62 \pm 2$
        & $5\,183$ & $456 \pm 2$
        & -- & -- \\
        & C\,IV
        & $6\,616$ & $197 \pm 12$
        & $6\,610$ & $611 \pm 7$
        &  -- & $\gtrsim 6.3 \times 10^6$ \\
           & Continuum & -- & -- & --  & -- & $93 \pm 4$ &--\\[2pt]
        Image C & Ly$\alpha$
        & $5\,184$ & $63 \pm 3$
        & $5\,182$ & $472 \pm 5$
        & -- & -- \\
        & C\,IV
        & $6\,615$ & $74 \pm 16$
        & $6\,613$ & $599 \pm 18$
        &  -- & $\gtrsim 4.8 \times 10^6$ \\
           & Continuum & -- & -- & --  & -- & $62 \pm 4$ &--\\
        \hline
    \end{tabular}%
    }
\end{table*}

\subsubsection{Eddington ratios}
\label{subsubsec:sec_edd-ratios}

Combining the SMBH mass lower limits for images A--C presented above with the strong lensing magnifications for images A \& B and image C ($\mu_{A} \sim  2$, $\mu_{B} \sim 2$ and $\mu_{C} \sim 350$) from the two-source lens model of \cite{Spingola2019}, we proceed to estimate the Eddington ratios of images A--C.~These are defined as:
\begin{equation}
\label{eq:eq1}
\lambda_{\text{Edd}} = \frac{L_{\text{bol,intrinsic}}}{L_{\text{Edd}}},
\end{equation}

\noindent where $L_{\text{bol, intrinsic}}$ is the intrinsic, magnification-corrected AGN bolometric luminosity and $L_{\text{Edd}}$ is Eddington luminosity, given by:

\begin{equation}
\begin{split}
    L_{\text{Edd}} & \approx {1.26\times 10^{38}} \left(\frac{M_{\text{BH}}}{M_\odot}\right) \, \text{erg s}^{-1}.\\ 
\end{split}
\label{eq:eq2}
\end{equation}

\noindent We calculate the intrinsic bolometric luminosity $L_{\text{bol, intrinsic}}$ for each lensed image from its corresponding observed bolometric luminosity ($L_\mathrm{bol, obs}$), following:

\begin{equation}
\label{eq:eq3}
    L_{\text{bol, intrinsic}} = \frac{L_{\text{bol, obs}}}{\mu_{\text{}}} = \frac{4\pi D_L^2 \, C_{\text{bol}} \sum_{i=1}^{N} F_i}{\mu_{\text{}}},
\end{equation}

\noindent where $\sum_{i=1}^{N} F_i$ is the total observed flux summed across all pixels in a given lensed image ($i=1,...,N$), $\mu_{\text{}}$ is the magnification factor \citep[which we take from][]{Spingola2019}, $D_L$ is the luminosity distance to the AGN redshift of $z=$~\sourceredshift, and $C_{\text{bol}}$ is the bolometric correction factor applied to observed continuum flux.~For the monochromatic continuum luminosity at rest-frame 1\,350\ \AA, the intrinsic bolometric luminosity can be estimated as:

\begin{equation}
    L_{\text{bol, intrinsic}} = \frac{C_{\text{1350}} \lambda L_{\lambda}(1\,350\ \mathring{\rm A})}{\mu} 
    =  \frac{4\pi D_L^2 \, C_{\text{1350}} \; \lambda f_{1350}}{\mu_{\text{}}} 
\label{eq:eq4}
\end{equation}
\noindent where the bolometric correction factor is $C_{\text{1350}}\approx 3.81$ and is typically used on quasars~\citep[][]{2006AJ....131.2766R,2011ApJS..194...45S}.

Following the dependence of both the SMBH mass and the intrinsic bolometric luminosity on the observed 1\,350\AA\ continuum flux (Equations \ref{eq:eq0} and \ref{eq:eq4}, respectively),~the Eddington ratio (Equations~\ref{eq:eq2}~and~\ref{eq:eq3})~follows the proportionality relation $\lambda_\mathrm{Edd} \propto {[\lambda f_{1350}]}^{0.47}$.~Along the sightlines of images A--C, the proximate absorbers at $z\sim3.27$ and the absorbers at $z\sim 0$ (described in Section \ref{sec:sec3_extraction})~attenuate the observed flux at rest-frame 1\,350\AA.~Therefore,~the absorption-uncorrected continuum flux densities in Table \ref{tab:fitted_parameters}~will lead to underestimated SMBH masses (as described in Section \ref{subsubsec:sec_smbh-masses}),~underestimated bolometric intrinsic luminosities,~and thus underestimated Eddington ratios.~Following the above formulae~(Equations \ref{eq:eq1} -- \ref{eq:eq4}),~we infer the following approximate Eddington ratios for images A--C:~$L_{\mathrm{bol,A}}/L_{\mathrm{Edd,A}}\approx 0.5$, $L_{\mathrm{bol,B}}/L_{\mathrm{Edd,B}}\approx 0.5$, and $L_{\mathrm{bol,C}}/L_{\mathrm{Edd,C}}\approx 0.04$.~These values should nevertheless be interpreted with additional caution, especially since image C is subject to differential magnification \citep[][]{Spingola2019} and $\mu_\mathrm{C}\sim350$ represents an average magnification factor.~Furthermore, while $\mu_\mathrm{C}\sim350$ is an average magnification factor that affects the radio VLBI source 1 component, its corresponding source-plane region probed in the \muse\ FOV may be spatially distinct and may therefore be subject to a different strong lensing magnification.

To provide an independent check on the Eddington ratio estimates inferred from the \muse\ data,~we also estimate the Eddington ratios using the X-ray luminosities of sources 1 and 2 inferred from the new \axaf\ observations of \sourceshortname\ presented in \citet{schwartz_mgb2016_2026}.~Following their section 3, we estimate their unobscured, intrinsic rest-frame 2--10 keV luminosities by extrapolating from the observed 2--30 keV luminosities using their best-fitting spectral models, obtaining $L_{\mathrm{2-10\ keV}}\approx 1.0\times10^{43}\ \mathrm{erg~s^{-1}}$ and $L_{\mathrm{2-10 \ keV}}\approx 4.3\times10^{43}\ \mathrm{erg~s^{-1}}$ for sources 1 and 2,~respectively. Adopting a representative bolometric correction of $\kappa_{\mathrm{2-10}}\approx30$ for high-$z$ quasars \citep{runnoe_2012_corrections} and our SMBH mass lower limits yields the following Eddington ratio estimates: $L_{\mathrm{bol,A}}/L_{\mathrm{Edd,A}}\approx 0.9$, $L_{\mathrm{bol,B}}/L_{\mathrm{Edd,B}}\approx 1.6$, and $L_{\mathrm{bol,C}}/L_{\mathrm{Edd,C}}\approx0.5$.~These X-ray-based estimates should be interpreted with caution, however, as the 2--10 keV bolometric correction can exhibit substantial scatter \citep{vasudevan_2007,vasudevan_2009},~which may be even more pronounced in high-redshift sources \citep[as recently demonstrated in the $z \approx 2 - 4$ sample of][]{Degli-Agosti_2025_kbols_high-z}. Thus, rather than providing precise measurements of the Eddington ratios, these X-ray estimates serve primarily as an independent consistency check on the relative accretion rates inferred from the \muse\ data.~In particular, despite the uncertainties associated with the X-ray bolometric correction and the fact that our SMBH masses are only lower limits,~they independently support the conclusion that the SMBH associated with source 1 is accreting at a lower rate than that associated with source 2.

If representative of the intrinsic accretion rates, our Eddington ratio estimates from the \muse\ data suggest that the lower-mass companion (associated with source 1) is accreting at a lower fraction of its Eddington limit than the more massive companion associated with source 2. Therefore, these limits may be of interest in the context of cosmological simulations, which often predict preferentially enhanced accretion onto the lower-mass companion in dual SMBH systems \citep{Steinborn2016,Volonteri2022,chen_2023_astrid}.

\section{Conclusions}
\label{sec:sec5_conclusions}

\begin{itemize}

\item We have presented archival \muse\ observations of the three spatially resolved, optical lensed images of the radio-loud system \sourcefullname\ (Figures \ref{fig:figure1_hst+muse} and \ref{fig:figure2_muse}).~At a projected separation of \offset\ at $z =$~\sourceredshift, this system represents the closest-separation confirmed dual AGN identified to date at such a high redshift \citep{Spingola2019,schwartz_mgb2016_2026}.
\item Previous optical spectroscopic analyses of \sourceshortname\ were focused on detecting and comparing various emission lines in images A and B \citep{schneider_1985_paperii,Schneider1986},~and in images B and C \citep{Yamada2001}.~The \cfht\ study of \citet{Yamada2001} reported discrepant \heliumtwo/\carbonfour\ and \carbonthree/\carbonfour\ ratios for images B and C, which were interpreted as arising from an ionization gradient across the narrow-line region of a single, type-II lensed AGN \citep{Yamada2001}.

\item In our \muse\ analysis,~we extracted emission spectra for the three lensed images A--C using 5-pixel apertures in radius around each image after correcting for ISM absorption at $z\sim0$.~Consistent with earlier spectroscopic studies \citep{schneider_1985_paperii,Schneider1986,Yamada2001}, we detect \lyalpha, \carbonfour, \nitrogenfive, \heliumtwo, and \carbonthree\ emission at high statistical significance in all three lensed images (Figure \ref{fig:figure3_spec}).~While the \lyalpha, \carbonfour, and \nitrogenfive\ emission lines are strongly affected by saturated absorption, the \carbonthree\ and \heliumtwo\ lines remain unaffected.~The \heliumtwo/\carbonthree\ ratios in images A and B -- i.e the ratio between the two lines that are unaffected by saturated absorption -- are consistent with one another but differ significantly from that measured in image C (Table \ref{tab:line_ratio_comparison}).~This discrepancy provides strong, independent spectroscopic evidence for two distinct AGN in~\sourcefullname~\citep[following the recent confirmation of its dual AGN nature from \axaf\ observations by][]{schwartz_mgb2016_2026}.

\item We fitted the \lyalpha~and \carbonfour~emission lines in images A--C using two Gaussian components where warranted by the data (Figure \ref{fig:spectral-fitting}; Table \ref{tab:fitted_parameters}).~All emission regions are well described by two Gaussians (with each Gaussian capturing a broad or narrow component).~For the first time, we use the broad Gaussian component of \carbonfour\ to place lower limits on the SMBH masses associated with the distinct lensed images (Table \ref{tab:fitted_parameters}) using the broad-line virial scaling relation of \citet{2006ApJ...641..689V} and the 1\,350\AA\ integrated luminosities found for each lensed image.~We place lower limits to the SMBH masses associated with images A--C:~$M_\mathrm{SMBH,A} \gtrsim$~\massalowlim, $M_\mathrm{SMBH,B} \gtrsim$~\massblowlim,~and $M_\mathrm{SMBH,C} \gtrsim$~\massclowlim.

\item Combining our SMBH mass estimates with the average strong lensing magnifications $\mu$ of the two-source model of \cite{Spingola2019}~and the integrated (observed) 1\,350\AA\ luminosities of images A--C, we have provided the first estimates of their corresponding Eddington ratios:~$L_{\mathrm{bol,A}}/L_{\mathrm{Edd,A}} \approx 0.47$, $L_{\mathrm{bol,B}}/L_{\mathrm{Edd,B}}\approx 0.51$, and $L_{\mathrm{bol,C}}/L_{\mathrm{Edd,C}}\approx 0.04$.~These estimates are consistent with the Eddington ratio upper limits derived independently from the rest-frame 2--10 keV X-ray luminosities for sources 1 and 2 from \cite{schwartz_mgb2016_2026}.~If the Eddington ratios we have estimated from \muse\ are indeed representative of the accretion state of the two SMBHs in \sourceshortname,~these would suggest that the high-mass SMBH is accreting at a higher rate than the lower-mass companion -- itself accreting below the quasar rate.~In such a scenario,~this dual AGN would occupy an interesting region of parameter space for dual AGN.~State-of-the-art simulations preferentially fuel the lower-mass system in compact SMBH pairs at a higher Eddington ratio.

\item In line with recent \muse-enabled studies of dual and lensed AGN \citep{scalpi_dualAGN_2026} and galaxies lensed by clusters \citep{claeyssens_2022_llamas}, this work highlights the potential of \muse\ to uncover complex absorption and emission in distant, gravitationally lensed dual AGN systems.~The strong saturation affecting several of the rest-frame UV emission lines, together with the lower limits on the SMBH masses we derived from our \carbonfour~emission-region fits, motivates deep, spatially resolved near-IR spectroscopy to obtain rest-frame optical diagnostics~(e.g.~\halpha, \hbeta)~that are unaffected by saturated absorption and can be used to independently revisit our black hole mass estimates.~Such observations are also critical to further constrain the ionization, kinematics, and physical conditions of the gas surrounding the dual AGN in \sourceshortname\ whilst shedding light on the host galaxy environment.

\end{itemize}

\begin{acknowledgments}
We are grateful for the referee's insightful and constructive comments, which have helped to improve the quality of this paper.~We are also grateful to Cristiana Spingola for constructive comments.~JH and VPK acknowledge partial support from the National Science Foundation grant AST/2009811 (PI:~Kulkarni).~Additional support from NASA/STScI grant HST-GO-17711.002 (PI:~De Cia) is greatly appreciated.~JSR acknowledges support from a NASA ADAP Program Grant 80NSSC24K0617.~DAS was supported by the National Aeronautics and Space Administration through the contract NAS8-03060 to SAO and grant GO4-25053X from the \axaf\ X-Ray Center (CXC).~CE thanks Daniel Castro for his support through grant GO1-22074X from the CXC.

This work has employed data from the \textit{Hubble Space Telescope} through the Mikulski Archive for Space Telescopes (MAST).~The Space Telescope Science Institute is operated by the Association of Universities for Research in Astronomy, Inc., under NASA contract NAS 5-26555.

This research has made use of the Astrophysics Data System, funded by NASA under Cooperative Agreement 80NSSC21M0056 and is now the Science Explorer, funded by NASA under Cooperative Agreement 80NSSC21M00561.

\textit{Facilities}:~HST, VLT/MUSE, VLBA. 

\textit{Software}:~\texttt{ds9}~v.8.7 \citep{ds9-joye-2003,fruscione_2026_recent}; Astropy \citep{astropy_2013,astropy_2018,astropy-collab-22}, and SciPy \citep{scipy-release-2020-virtanen}.~Astropy and Scipy were used to compute the observed fluxes for images A--C and to fit their emission-line spectra with two Gaussians.

\section*{Author contributions}

This work was co-led by JH and JSR.~JH performed the data reduction and analysis of the \muse\ data and contributed significantly to the interpretation of the results.~JSR wrote the manuscript,~coordinated our team's meetings,~and contributed significantly to the interpretation of the results.~JSR and DS led the interpretation of the \axaf\ data discussed in the text.~DS also contributed significantly to the interpretation of the results and provided comments on the manuscript.~VK contributed to the spectroscopic interpretation of the absorbers inferred from the \muse\ data and to scientific discussions, and provided comments on the manuscript.~AB, AF, and CE provided comments on the manuscript and contributed to scientific discussions.~AF contributed to the interpretation of the SMBH limits inferred from the \muse\ data and contributed to its discussion.~CE produced the \hst\ image shown in Figure \ref{fig:figure1_hst+muse}.

\end{acknowledgments}

\bibliography{sample701}{}
\bibliographystyle{aasjournalv7}

\end{document}